\documentclass[%
reprint,
amsmath,amssymb,
aps,
]{revtex4-2}

\usepackage{graphicx}% Include figure files
\usepackage{dcolumn}% Align table columns on decimal point
\usepackage{bm}% bold math
\usepackage{color}

\begin{document}
	
	%\preprint{APS/123-QED}
	
	%\title{Rate dependence of 3D DDD}% Force line breaks with \\
	\title{Avalanches and rate effects in strain-controlled\\ discrete dislocation plasticity of Al single crystals}
	
	\author{David Kurunczi-Papp}
	\email{david.kurunczi-papp@tuni.fi}
	\author{Lasse Laurson}%
	\affiliation{%
		Computational Physics Laboratory, Tampere University, P.O. Box 692, FI-33014 Tampere, Finland
	}%
	
	\date{\today}% It is always \today, today,
	
	\begin{abstract}
		Three-dimensional discrete dislocation dynamics simulations are used to study strain-controlled plastic deformation of face-centered cubic aluminium single crystals. %crystalline solids. 
		After describing the rate and size dependence of the average stress-strain curves, we study the power-law distributed strain bursts, and find a universal power-law exponent $\tau \approx 1.0$ for all imposed strain rates. This is then followed by the characterization of the average avalanche shapes which reveals the two key physical regimes in dislocation plasticity dominant at small and large strain rates, respectively. We discuss the dependence on the loading rate and compare our observations with previous studies of strain-controlled two-dimensional systems of discrete dislocations as well as of quasistatic stress-controlled loading of aluminium single crystals.
	\end{abstract}
	
	\maketitle
	
	%\tableofcontents
	
	\section{\label{sec:1} Introduction}
	
	The initial elastic deformation of crystalline solids subject to external stresses undergoes a continuous transition ("yielding") to plastic flow which is governed by the collective dynamics of dislocations~\cite{papanikolaou2017avalanches,alava2014crackling}. This complex motion of dislocation systems is characteristic of strongly intermittent, avalanchelike fluctuations exhibiting broad, power-law-like size distributions~\cite{csikor2007dislocation,Dimiduk1188}. These were observed by both acoustic emission measurements on macroscopic samples~\cite{weiss2003three,PhysRevLett.114.105504} and in experiments performed on micron-scale samples~\cite{PhysRevB.79.014108,SALMAN2012219,PhysRevE.102.023006,alcala2020}. Despite all the recent advances, the dependence of dislocation plasticity on the crystal structure and orientation as well as size- and rate effects of the deformation process lacks a complete description~\cite{AGNIHOTRI201537,PhysRevMaterials.2.120601,fan2021strain}. Moreover, the origin of the bursty nature of crystal plasticity is not yet fully understood~\cite{PhysRevLett.109.095507,PhysRevE.91.042403,ispanovity2014avalanches}, although avalanche mechanisms were successfully described using nanopillar compression models~\cite{CROSBY2015123,PhysRevB.95.064103,PhysRevMaterials.3.080601,CUI2020117}.
	
	Discrete dislocation dynamics (DDD) simulations have been proven to be efficient numerical tools in the study of crystal plasticity on the microscopic scale, capturing the avalanchelike deformation process~\cite{PhysRevLett.89.255508,csikor2007dislocation,ispanovity2014avalanches,ovaska2015quenched}. In the simplest and computationally most efficient DDD models the dislocations are represented as point-like objects (representing cross sections of straight parallel edge dislocations) moving and interacting within a single plane~\cite{PhysRevLett.89.165501,PhysRevLett.105.015501,PhysRevE.104.025008}. More realistic 3D systems describing dislocations as flexible lines, including edge, screw as well as mixed dislocations, are able to provide a detailed insight into the complex nature of dislocation systems, taking into consideration the different crystal structures and dislocation movements along multiple glide planes~\cite{ParaDiS,lehtinen2016glassy}. An additional advantage over 2D DDD simulations is the possibility to include the topological operations responsible for the increase of the dislocation density during the deformation process.
	
	Previous studies have mainly employed a quasistatic stress-controlled loading, both in 2D models~\cite{ispanovity2014avalanches,Salmenjoki2018} and in 3D simulations of FCC aluminium single crystals~\cite{lehtinen2016glassy,PhysRevMaterials.5.073601}, where the stress is either increased constantly as long as the collective dislocation velocity is below a threshold, or is held constant with accumulating strain (avalanche) until this velocity relaxes to a value below the threshold. Quasistatic stress-controlled loading simulates the low stress rate limiting behavior of the deformation process, eliminating possible rate effects in the avalanche statistics. However, strain-controlled loading allows simulations with strain rates spanning up to nine orders of magnitudes including both forest hardening and strain rate hardening regimes, as it was demonstrated for the material parameters of FCC copper~\cite{fan2021strain}. In this paper, we present results from an extensive study of 3D DDD using strain-controlled loading of FCC aluminium single crystals. sing a large database of stress-strain curves, we characterize the size- and rate effects of the ensemble averages, and we show that the avalanches are distributed according to the same power-law, independent of the deformation rate. On the other hand, the avalanche shapes reveal a rate dependence, separating the deformation process into two regimes dominated by different mechanisms. Alongside the analysis, we compare our observations with the results from related studies on 2D systems~\cite{PhysRevE.104.025008} as well as on stress-controlled loading of 3D systems~\cite{lehtinen2016glassy}.
	
	The paper is organized as follows: In Sec.~\ref{sec:2} we present the DDD model used and illustrate a sample deformation process. The results of our study are presented in Sec.~\ref{sec:3}, starting with the size- and rate effects of the stress-strain curves in Sec.~\ref{subsec:3.1}, followed by the statistical analysis of the avalanche distributions, scaling as a function of their duration, and average shapes in Sec.~\ref{subsec:3.2}. The conclusions of our paper are summarized in Sec.~\ref{sec:4}.
	
	\section{\label{sec:2} DDD Simulations}
	
	To simulate the deformation process, we use the constant strain rate loading method of the 3D DDD software ParaDiS \cite{ParaDiS}. Here the dislocation lines are discretized into nodal points connected by straight line segments. Dislocation motion is realized by moving the nodal points, which, additionally, can be added or removed depending on segment lengths and curvatures. The total stress acting on a node is the sum of the external stress, generating a Peach-Koehler force, and of the anisotropic stress fields generated by the dislocations within the crystal~\cite{Dislocations}. The forces generated by the latter fields are divided into local and far-field ones. The local forces are computed directly via line integrals, while the far-field forces are obtained from the course-grained dislocation structure using a fast multipole method (FMM)~\cite{greengard_rokhlin_1997}. FMM works without the need of a cut-off distance and is widely used in other physical systems that contain long range forces, such as Coulomb interactions in ionic crystals or gravitational fields in galaxies. The calculation of the forces is followed by the time integration and solving the equations of motion for the discretized nodes. ParaDiS takes into account the crystal structures and dislocation types (edge, screw, or a combination of those). These material specific properties enter the equations of motion through the mobility functions, relating the forces with the dislocation velocities. Bulk properties are simulated by using periodic boundary conditions (PBC). The interactions of the dislocation segments in the simulation box with all of their periodic images is handled within the FMM algorithm.
	
	We consider here the FCC crystal structure with material parameters of Al: shear modulus $G=26\,\mathrm{GPa}$, Poisson ratio $\nu=0.35$, Young modulus $Y=70.2\,\mathrm{GPa}$, Burgers vector $b=2.863\times10^{-10}\,\mathrm{m}$ and, for simplicity, the same dislocation mobility $M=10^4\ \mathrm{Pa}^{-1}s^{-1}$ in all directions. To study the effect of the system size, we consider different linear sizes $L=0.741$, $1.054$ and $1.43\,\mathrm{\mu m}$ of the cubic simulation box (i.e. within the range of those of typical microcrystal compression experiments~\cite{Dimiduk1188}). These system sizes are chosen in a way that the numbers $N_0=10$, $20$ and $40$ of the initially straight dislocation lines placed randomly on the glide planes of the FCC lattice, relax in zero applied stress to a dislocation configuration in a metastable state with the approximate density $\rho_0\approx2.5\times10^{13}\,\mathrm{m}^{-2}$. After the initial relaxation stage, the strain-controlled loading protocol (constant strain rate loading method of ParaDiS) is switched on imposing strain rates from $\dot{\epsilon}_a=1000\,\mathrm{s}^{-1}$ up to $\dot{\epsilon}_a=2\times10^5\,\mathrm{s}^{-1}$. For the $N_0=10$ and $40$ system sizes $100$ and $50$ samples were generated for every imposed strain rate, respectively, while for the medium system size $N_0=20$ (used to study the avalanche statistics) $100$ and $200$ samples for the two lowest ($\dot{\epsilon}_a=1000\,\mathrm{s}^{-1},\ 5000\,\mathrm{s}^{-1}$) and the higher rates (above $\dot{\epsilon}_a=10000\,\mathrm{s}^{-1}$) were created, respectively. The resulting stresses $\sigma$, total strains $\epsilon$ as well as plastic strains $\epsilon_{\mathrm{P}}$, loading times $t$, strain rates $\dot{\epsilon}$ and dislocation densities $\rho$ are stored during the simulations.
	
	\begin{figure}[h!]
		\centering
		\resizebox{0.95\columnwidth}{!}{\includegraphics{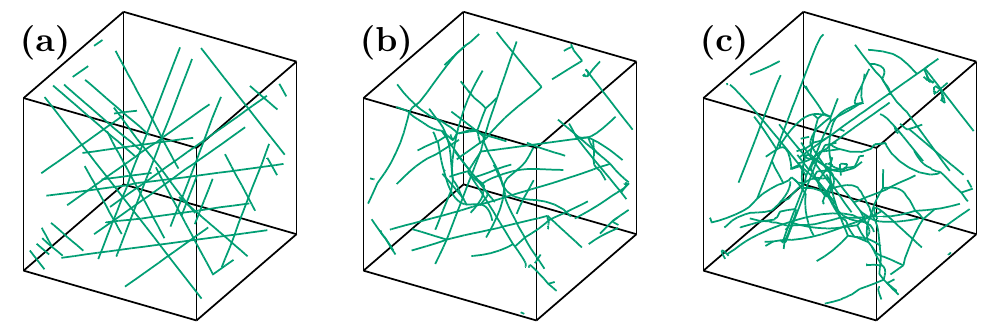}}
		\resizebox{0.95\columnwidth}{!}{\includegraphics{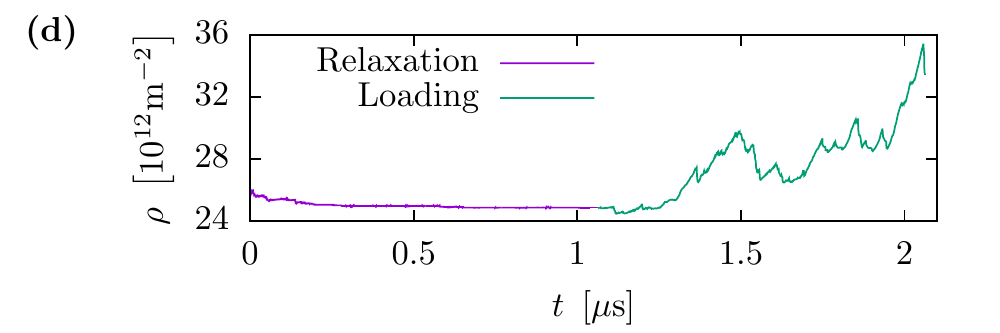}}
		\resizebox{0.95\columnwidth}{!}{\includegraphics{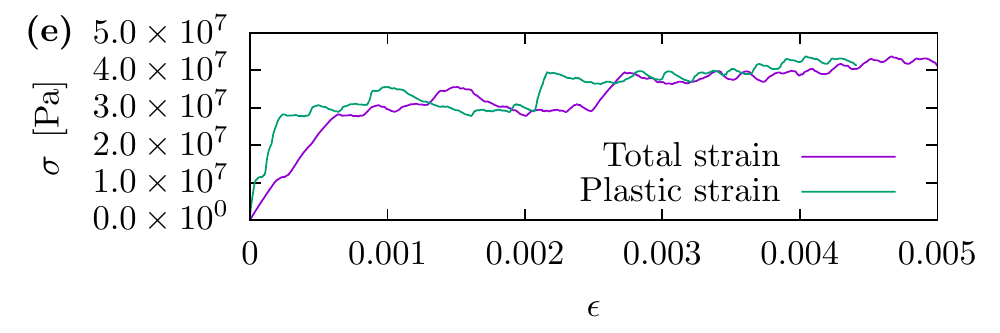}}
		\caption{Sample dislocation configurations of the system with $N_0=20$ initially straight dislocations: (a) before relaxation, (b) after relaxation and (c) deformed up to $\epsilon=0.005$ with a strain rate $\dot{\epsilon}_r=5000\,\mathrm{s}^{-1}$; (d) the evolution of the dislocation density during the relaxation as well as the loading process, and (e) the resulting stress-strain curve, illustrating the difference between the plastic and total strains.}
		\label{fig:1}
	\end{figure}

	A sample simulation process with $N_0=20$, $\dot{\epsilon}_a=5000\,\mathrm{s}^{-1}$ is illustrated in Fig.~\ref{fig:1}, including the straight dislocations placed randomly in the simulation box, the relaxed configuration where the dislocations exhibit some curvature, the configuration at the end of the loading procedure, the evolution of the dislocation density throughout the whole process, and the resulting stress-strain curves. The last panel shows the stress as a function of both the plastic and total strain. At the beginning of the deformation process the total strain is dominated by its elastic component, in accordance with linear elastic theory. This is then followed by the plastic deformation, exhibiting a sawtooth-like shape, consisting of a series of stress drops or strain bursts and segments with increasing $\sigma$.

	\section{\label{sec:3} Results}
	
	\subsection{\label{subsec:3.1} Rate-dependent stress-strain curves}
	
\begin{figure}[h!]
		\centering
		\resizebox{0.95\columnwidth}{!}{\includegraphics{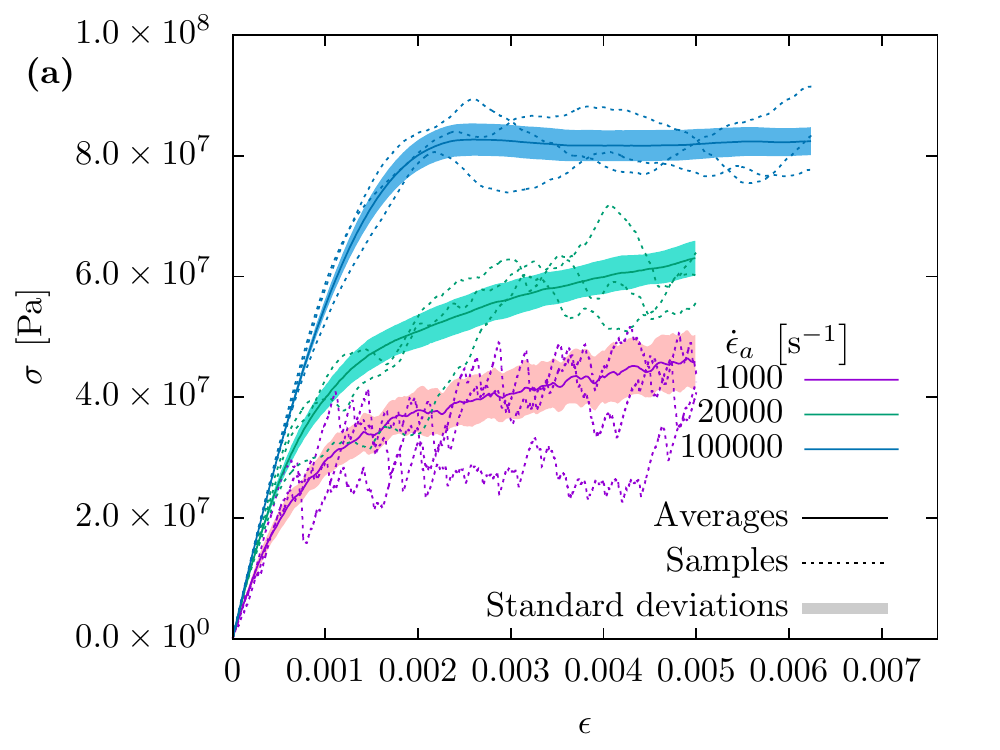}}
		\resizebox{0.95\columnwidth}{!}{\includegraphics{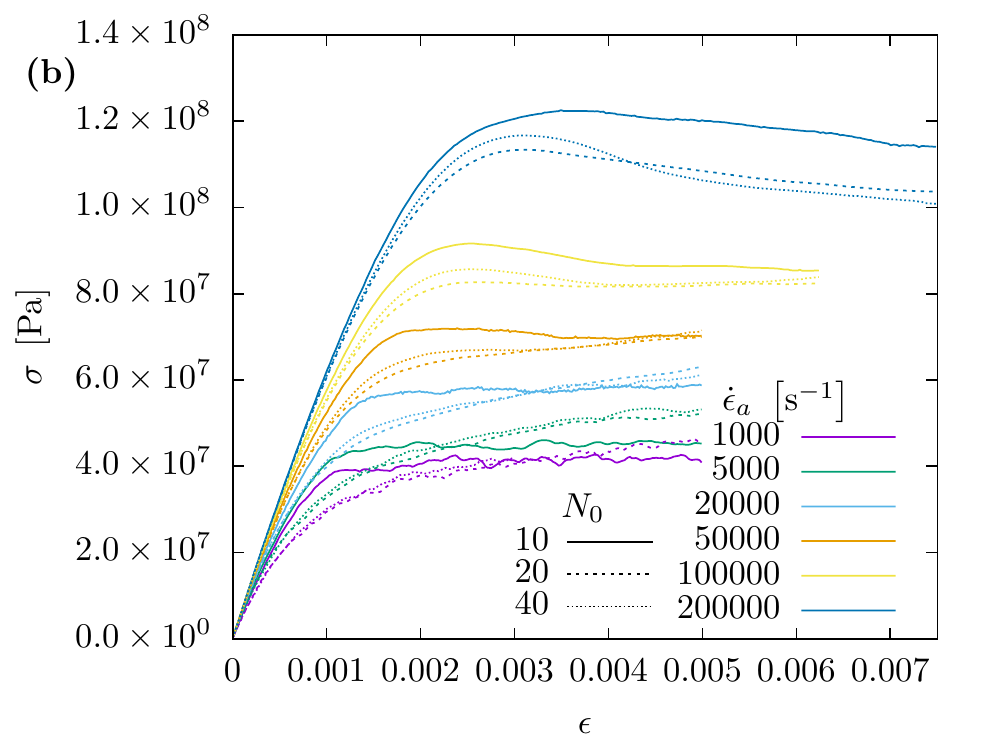}}
		\caption{(a) Average stress-strain curves $\langle \sigma(\epsilon)\rangle$ (with the average taken over different realizations of the initial configuration, solid lines), individual, single sample curves (dashed lines), and standard deviations (shaded regions) for system with initial dislocation lines $N_0=20$ and various applied strain rates $\dot{\epsilon}_a$. (b) Average stress-strain curves $\langle \sigma(\epsilon)\rangle$ (with the average taken over different realizations of the initial configuration) for varying applied strain rates $\dot{\epsilon}_a$ and three system sizes defined by the number of initial dislocation lines: $N_0=10$ solid lines; $N_0=20$ dashed lines; $N_0=40$ dotted lines.}
		\label{fig:2}
	\end{figure}

	First we investigate the rate and size dependence of the stress-strain curves. Figure~\ref{fig:2}(a) shows examples of single-sample curves as a function of the total strain along with the ensemble-averaged stress-strain curve and its standard deviation for the system with $N_0=20$ initial dislocation lines and imposed strain rates $\dot{\epsilon}_a$ in three different orders of magnitude. Varying the driving parameter $\dot{\epsilon}_a$ results in a significant dependence of the fluctuation magnitudes and average stress-strain curves on $\dot{\epsilon}_a$. The system size dependence along with the rate dependence of the ensemble-averaged stress-strain curves is visualized in Fig.~\ref{fig:2}(b). While the rate dependence highly influences the stresses reached during the loading process, a non-negligible size dependence is also present in the plastic regime. The smallest systems with $N_0=10$ initial dislocation lines no longer accumulate further stress past the elastic regime, and the average stress-strain curves approach a fixed stress value, provided there are sufficient samples to average out the fluctuations, as in the case of larger rates. Meanwhile, the average stress-strain curves for the larger systems and smaller rates keep increasing monotonically beyond the initial elastic deformation, transitioning to a monotonically decreasing averaged stress-strain curve around the imposed strain rate $\dot{\epsilon}_a=100000\,\mathrm{s}^{-1}$. Due to this similar behavior for $N_0=20$ and $40$, as well as the high computational cost of larger systems, the remainder of this work will focus on the systems with $N_0=20$ initial dislocation lines placed in the periodic simulation box.
	
\begin{figure}[h!]
		\centering
		\resizebox{0.95\columnwidth}{!}{\includegraphics{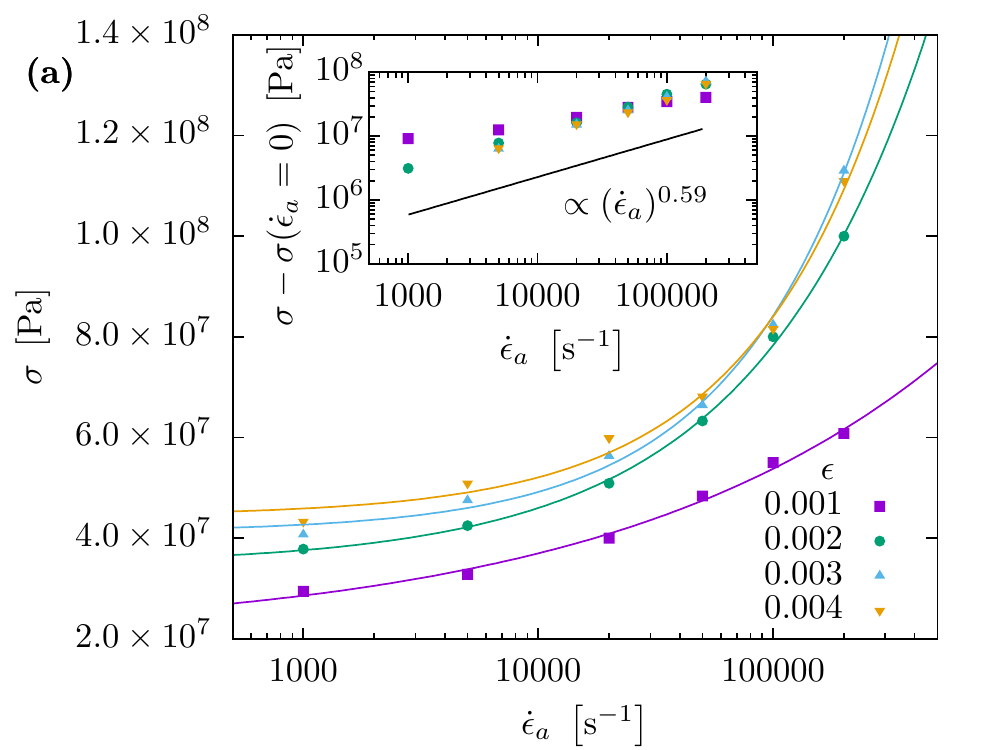}}
		\resizebox{0.95\columnwidth}{!}{\includegraphics{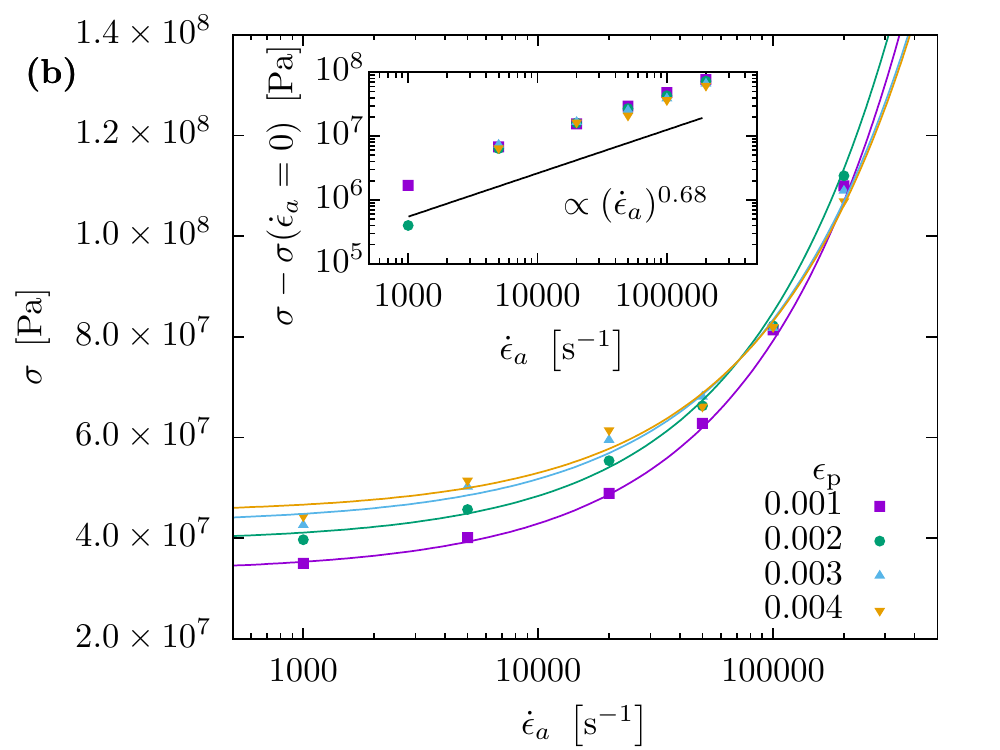}}
		\caption{The dependence of the ensemble-averaged stress-strain curves on the strain rate $\dot{\epsilon}_a$ at specific (a) strain $\epsilon$ and (b) plastic strain $\epsilon_{\mathrm{p}}$ values, fitted to a power-law shifted by the stress at zero rate given by Eq.~\eqref{eq:1}. The insets show the stress with the zero rate values subtracted on a logarithmic scale and the corresponding power laws (a) $b=0.59$ and (b) $b=0.68$ for the strain and plastic strain values, respectively.}
		\label{fig:3}
	\end{figure}

	The dependence of the accumulated average stress at given strain as a function of the imposed strain rate is shown in Fig.~\ref{fig:3}. We expect the dependence to be described by a shifted power law 
	\begin{equation}\label{eq:1}
	    \sigma(\epsilon,\dot{\epsilon}_a)= \sigma(\epsilon,\dot{\epsilon}_a=0)+A(\dot{\epsilon}_a)^b
	\end{equation}
	with the shift equal to the stress at zero strain rate~\cite{PhysRevE.104.025008}. Figure~\ref{fig:3}(a) considers the total strain containing both the elastic and plastic strains. The lowest strain considered $\epsilon=0.001$ is dominated by the elastic component and thus we observe the low exponent $b\approx0.3$. Meanwhile the strain dominated by the plastic strains obey the power law $b\approx0.59$. On the other hand, considering only the plastic strain [see Fig.~\ref{fig:3}(b)] gives us the higher exponent $b\approx0.68$ for all the plastic strain values tested. Both values obtained for the deformation dominated by plastic strains are higher than the exponent $b\approx0.4$ found recently for 2D DDD simulations considering the point-like cross sections of parallel, straight edge dislocations \cite{PhysRevE.104.025008}. A similar trend was also observed for copper, where the exponent decreases with increasing dislocation density~\cite{fan2021strain}.
	
	\subsection{\label{subsec:3.2} Avalanche statistics}
	
\begin{figure}[h!]
		\centering
		\resizebox{0.95\columnwidth}{!}{\includegraphics{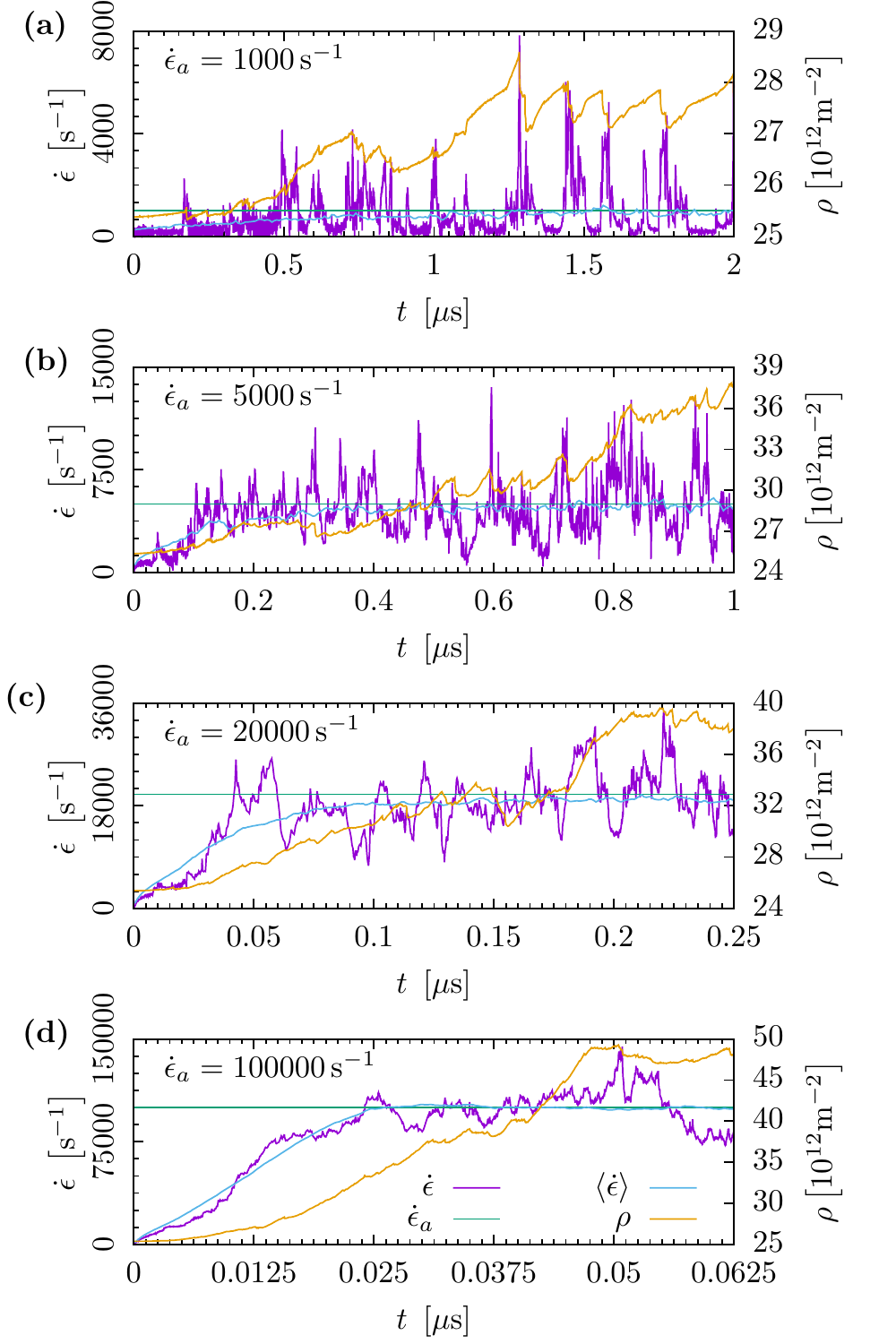}}
		\caption{Individual single sample strain rate signals $\dot{\epsilon}$, the applied strain rates (a) $\dot{\epsilon}_a=1000\,\mathrm{s}^{-1}$, (b) $\dot{\epsilon}_a=5000\,\mathrm{s}^{-1}$, (c) $\dot{\epsilon}_a=20000\,\mathrm{s}^{-1}$, (d) $\dot{\epsilon}_a=100000\,\mathrm{s}^{-1}$; the signal averages $\langle \dot{\epsilon} \rangle$ taken over different realizations of the initial configuration, and the corresponding single sample evolution of the dislocation densities for the system with $N_0=20$ initial dislocation lines.}
		\label{fig:4}
	\end{figure}

	In the following we study the dependence of the strain rate signal on the imposed strain rate. Figure~\ref{fig:4} illustrates how the strain rate signal $\dot{\epsilon}$, the ensemble averaged signal $\langle \dot{\epsilon} \rangle$, as well as the dislocation density $\rho$ evolves during the loading process for a wide range of imposed strain rates $\dot{\epsilon}_a$ (note the varying $y$-axis ranges). By definition of the loading mechanism, a signal above the threshold $\dot{\epsilon}>\dot{\epsilon}_a$ results in a monotonically decreasing stress rate $\dot{\sigma}$, thus creating a strain burst. This property of the signal allows us to better understand the fluctuating behavior of the stress-strain curves. The lowest imposed strain rate $\dot{\epsilon}_a=1000\,\mathrm{s}^{-1}$ [see Fig.~\ref{fig:4}(a)] results in a signal dominated by sharp peaks well above the threshold, while the signal below the threshold is close to zero. As the imposed strain rate is increased [see Fig.~\ref{fig:4}(b)], the fluctuations become more localized around the threshold, and the signal no longer approaches zero. These effects are enhanced by a further increase in the imposed strain rate [see Fig.~\ref{fig:4}(c)], as well as causing the fluctuations to be symmetric around the threshold, due to the higher gap above the zero rate. Figure~\ref{fig:4}(d) shows the strain rate signal for the high imposed strain rate $\dot{\epsilon}_a=100000\,\mathrm{s}^{-1}$, where the fluctuations around the threshold are small and the ensemble averaged signal $\langle \dot{\epsilon} \rangle$ overlaps well with the imposed strain rate, exhibiting minimal fluctuations. Similarly to the stress-strain curves, the dislocation densities are characterized by rate dependent fluctuations, where low- and high rates exhibit large and small fluctuations, respectively. Additionally, the comparison of the time evolutions of the strain rate signal and dislocation density reveals that at low rates the sharp peaks of the signal overlap well with the decreases in the dislocation density, while the same correlation becomes less prominent as the imposed strain rate approaches higher values. This hints towards the existence of two regimes in deformation of FCC single crystals: (i) at low rates the individual avalanches are separated in time, while (ii) at high rates several distinct avalanches could be propagating at the same time, resulting in a temporal overlap of the events. Here it should be noted that higher rates allow a deformation in smaller time scales, and thus the number of fluctuations during the loading process vary significantly for the imposed strain rates shown. This affects the avalanche statistics discussed in the following.

    %(i) at low rates the dislocations can respond in time to external changes meaning that only a few dislocations are moving at a time, while (ii) at high deformation rates the response to external changes stretches over a longer period of time, allowing the simultaneous activation of more dislocations.} Here it should be noted that higher rates allow a deformation in smaller time scales, and thus the number of fluctuations during the loading process vary significantly for the imposed strain rates shown. This affects the avalanche statistics discussed in the following.

\begin{figure}[h!]
		\centering
		\resizebox{0.95\columnwidth}{!}{\includegraphics{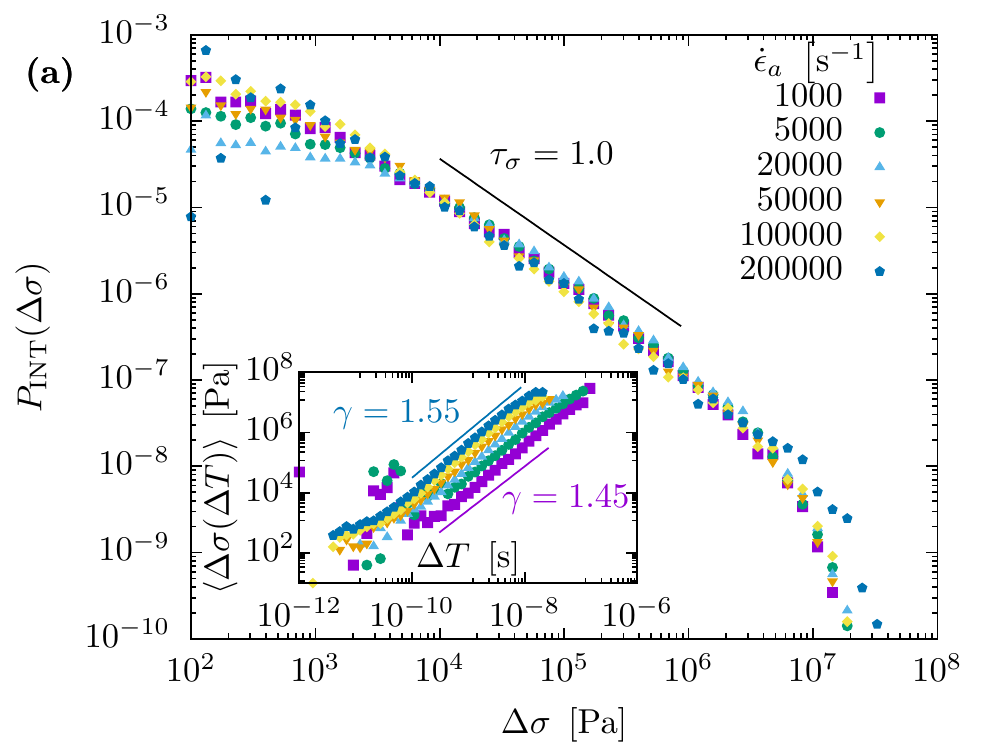}}
		\resizebox{0.95\columnwidth}{!}{\includegraphics{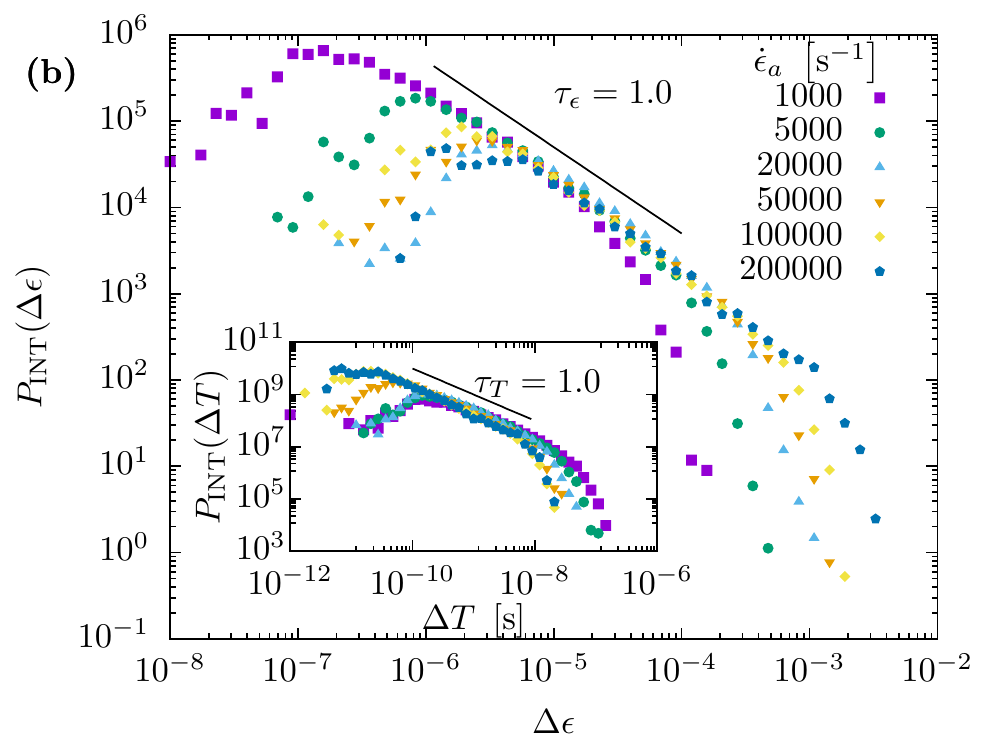}}
		\caption{(a) Integrated distributions of the stress drop magnitudes $P_{\mathrm{INT}}(\Delta \sigma)$ for varying imposed strain rates $\dot{\epsilon}_a$, described by the power law $\tau_\sigma=1.00\pm0.05$ with exponential cutoff. The inset shows the scaling of the average stress drop magnitude $\langle \Delta\sigma(T) \rangle$ with the event duration $T$, where the lines correspond to the power laws obtained for the lowest and highest $\dot{\epsilon}_a$, respectively. (b) Integrated distributions of the strain increments $P_{\mathrm{INT}}(\Delta \epsilon)$ for varying imposed strain rates $\dot{\epsilon}_a$, described by the power law $\tau_\epsilon=1.0\pm0.05$ with exponential cutoff. The inset shows the corresponding event duration distributions (proportional to the strain increments) obeying the same power law $\tau_T=1.0\pm0.05$.}
		\label{fig:5}
	\end{figure}

	Figure~\ref{fig:5}(a) shows the integrated stress drop distributions $P_{\mathrm{INT}}(\Delta \sigma)$, i.e., the distribution of the stress differences during avalanches irrespective of the stress- and strain value at which they occur. The distributions are fitted to a power law terminated at an exponential cutoff $P_{\mathrm{INT}}(\Delta \sigma) =A(\Delta\sigma)^{-\tau_\sigma}\exp{(-\frac{\Delta\sigma}{\Delta\sigma_0})}$, resulting in the universal exponent $\tau_\sigma=1.00\pm0.05$ independent of the imposed strain rate, and overlapping cutoff scales $\Delta\sigma_0$ up to an accuracy defined by the availability of statistics in the cutoff regime. The inset of Fig.~\ref{fig:5}(a) shows the scaling of the average avalanche size $\langle \Delta\sigma(\Delta T)\rangle$ with the avalanche duration $\Delta T$. The scaling is defined by the power $\gamma=1.5\pm0.05$ for the average avalanche size being in the power law regime. The seemingly small rate dependent variation exhibited by the scaling is within the error range, meaning that the rate dependence is only visible in the magnitudes of the stress drops of a given duration. This scaling relation is in accordance with the power law $\gamma=1.5\pm0.02$ obtained for large events in case of quasistatic stress-controlled loading of statistically equivalent Al single crystal samples~\cite{lehtinen2016glassy}.
	
	The corresponding integrated distributions of strain increments $P_{\mathrm{INT}}(\Delta \epsilon)$ as well as the avalanche durations $P_{\mathrm{INT}}(\Delta T)$ are visualized in the main panel and inset of Fig.~\ref{fig:5}, respectively. Similarly to the distributions of the stress drop magnitudes, the strain increments are distributed by a truncated power law with the exponent $\tau_\epsilon=1.00\pm0.05$ for all the tested imposed strain rates, however the scaling regime exhibits a rate dependence with an increase of the cutoff scale $\Delta\epsilon_0$ for increasing rates. Because the total strain $\epsilon$ is defined as the product of the imposed strain rate $\dot{\epsilon}_a$ and the elapsed time $t$, the avalanche duration distributions are shifted versions of the strain increment distributions, obeying the same power law $\tau_T=1.00\pm0.05$, and at the same time resulting in the reverse rate dependence, i.e a decrease of the cutoff scale $\Delta T_0$ for increasing rates.
	
	The avalanchelike dynamics of plastic deformation has been characterized by various power law exponents $\tau$. The value $\tau\approx1.5$ predicted by mean-field depinning~\cite{FISHER1998113} was found both experimentally and numerically~\cite{NG20081712,PhysRevLett.100.155502,Zaiser2008PhilosMag,lehtinen2016glassy}. However, the exponents found using 2D DDD simulations do not agree: quasistatic stress-controlled loading resulting in $\tau\approx1.0$ for stress resolved avalanche distributions (where the events are binned into stress intervals) and the larger exponent $\tau\approx1.3$ for the integrated distributions~\cite{ispanovity2014avalanches}, while avalanches from strain-controlled loading exhibit a large and rate dependent exponent~\cite{PhysRevE.104.025008}. Including the observation that the scaling of the average avalanche sizes $\langle \Delta\sigma(\Delta T)\rangle$ with the avalanche duration $\Delta T$ are described by the power laws $\gamma\approx1.0$ and $\gamma\approx1.5$ for two- and three dimensional DDD models, respectively, we can state that: (i) these two models belong to different universality classes of mesoscopic plasticity; and (ii) the different avalanche definitions for stress- and strain-controlled loading modes (constant and monotonically decreasing stress, respectively) affect the power-law exponent of the avalanche distributions. One possible explanation is provided by the choice of threshold strain rate (or velocity) above which avalanches are defined~\cite{lehtinen2016glassy,PhysRevE.104.025008,PhysRevE.105.054152}. Additionally, models simulating experiments (where pure stress or strain control cannot be achieved) of pillar compression have linked the transition between the two loading modes to the different microscopic mechanisms during sample deformation~\cite{PhysRevLett.117.155502,PhysRevB.95.064103}.
	
	%\begin{figure}[h!]
	%	\centering
	%	\resizebox{0.95\columnwidth}{!}{\includegraphics{time-stress-20.pdf}}
	%	\caption{Scaling of the average stress drop magnitude.}
	%	\label{fig:6}
	%\end{figure}
	
	Besides the avalanche sizes and durations, strain bursts can be characterized by their average shapes~\cite{Zapperi2005,PhysRevE.74.066106,Papanikolaou2011,laurson2013avalancheshape}. In plasticity the event shape means the average strain rate profile from the start of the strain burst to the end of it, with the imposed strain rate subtracted from the signal, which is proportional to the time derivative of the stress in that interval $\langle \dot{\sigma} (t/T) \rangle\propto\langle \dot{\epsilon} (t/T) - \dot{\epsilon}_a \rangle$. We expect such average shapes to be parameterized by the exponent $\gamma$, which we predict to be the same exponent characterizing the average stress drop magnitude scaling $\langle \Delta\sigma(\Delta T)\rangle$ with the event duration $\Delta T$, and the parameter $a$ describing the temporal asymmetry of the avalanches~\cite{laurson2013avalancheshape},
	
	\begin{eqnarray}\label{eq:2}
	{\textstyle \langle \dot{\sigma} (t/T) \rangle \propto T^{\gamma-1}\left[ \frac{t}{T}\left(1-\frac{t}{T}\right) \right]^{\gamma-1} \left[ 1-a\left(\frac{t}{T}-\frac{1}{2}\right) \right] }.
	\end{eqnarray}
	Eq.~~\ref{eq:2} consists of a symmetrical part parameterized by $\gamma$, and a lowest order correction to describe a weak asymmetry of the shape, quantified by $a$.
	
	\begin{figure}[h!]
		\centering
		\resizebox{0.95\columnwidth}{!}{\includegraphics{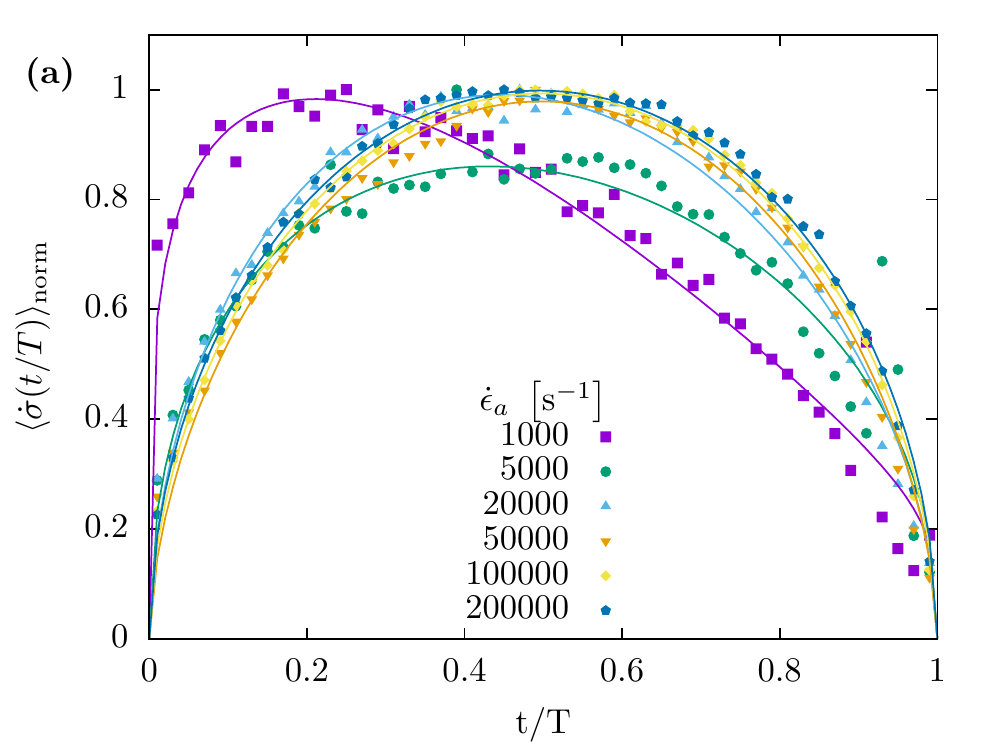}}
		\resizebox{0.95\columnwidth}{!}{\includegraphics{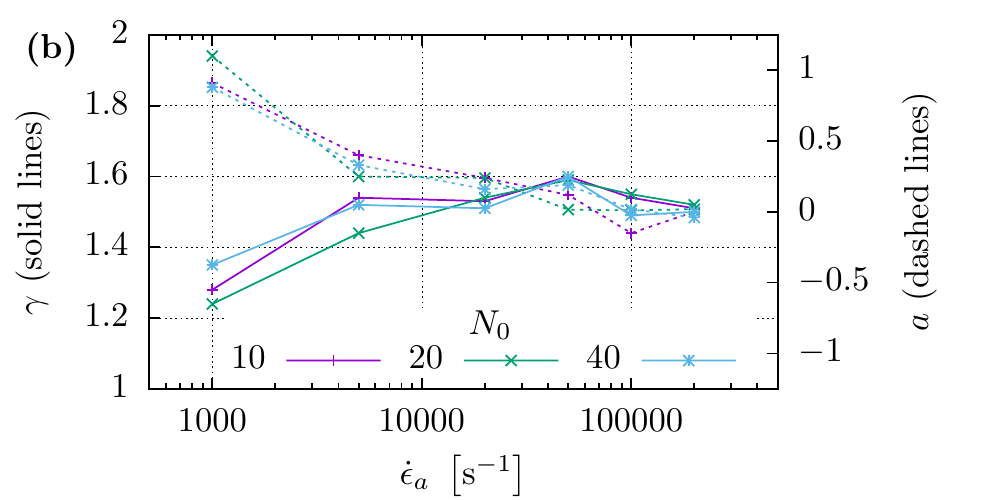}}
		\caption{(a) Average strain burst (avalanche) shapes normalized by their maximum values $\langle \dot{\sigma} (t/T) \rangle_{\mathrm{norm}}$ for events in the power-law regime of the size and duration distributions, for varying imposed strain rates $\dot{\epsilon}_a$, and for a system size with $N_0=20$ initial dislocation lines. (b) Size and rate dependence of the average strain burst shapes represented by the fitting parameters $\gamma$ and $a$ from Eq. \eqref{eq:2}.}
		\label{fig:6}
	\end{figure}

	\begin{table}[t]
	\caption{Exact $\gamma$ and $a$ parameter values used to represent the size and rate dependence of the average avalanche shapes in Fig.~\ref{fig:6}(b).}
	\label{tab:table1}
	\begin{tabular}{c|cc|cc|cc}
	%\begin{tabular}{p{0.1\columnwidth}>{\centering}p{0.15\columnwidth}>{\centering}p{0.25\columnwidth}>{\centering\arraybackslash}p{0.2\columnwidth}}
	    $N_0$    & \multicolumn{2}{c|}{$10$} & \multicolumn{2}{c|}{$20$} & \multicolumn{2}{c}{$40$} \\ \hline
		$\dot{\epsilon}_a\ \left[\mathrm{s}^{-1}\right]$   & $\gamma$ & $a$ & $\gamma$ & $a$ & $\gamma$ & $a$ \\ \hline
		$1000$   & $1.28$ &  $0.91$ & $1.24$ & $1.1$  & $1.35$ &  $0.88$ \\
		$5000$   & $1.54$ &  $0.4$  & $1.44$ & $0.25$ & $1.52$ &  $0.33$ \\
		$20000$  & $1.53$ &  $0.24$ & $1.54$ & $0.24$ & $1.51$ &  $0.16$ \\
		$50000$  & $1.6$  &  $0.12$ & $1.59$ & $0.02$ & $1.6$  &  $0.19$ \\
		$100000$ & $1.54$ & $-0.15$ & $1.55$ & $0.01$ & $1.49$ &  $0.02$ \\
		$200000$ & $1.51$ &  $0.0$  & $1.52$ & $0.02$ & $1.5$  & $-0.04$
	\end{tabular}
    \end{table}

	Figure~\ref{fig:6}(a) shows the average avalanche shapes for events with durations in the interval $\left[10^{-9}\,\mathrm{s};10^{-8}\,\mathrm{s}\right]$ found in the power-law regime of the size and duration distributions. The shapes are normalized by their maximal values within the interval. The shape for the lowest imposed strain rate $\dot{\epsilon}_a=1000\,\mathrm{s}^{-1}$ exhibits a high asymmetry, characteristic of sharp peaks caused by a jump of an individual dislocation, followed by a relaxation corresponding to a slow decay in the strain rate signal [see Fig.~\ref{fig:4}(a)]~\cite{Jani_evi__2015}. As the rate is increased, the average avalanches approach a symmetrical shape, a trend also observed in 2D DDD studies~\cite{PhysRevE.104.025008}. Our results can be related to the observations made for quasistatic stress-controlled loading of Al single crystals: small avalanches resulting in small strain, corresponding to low strain rates in our study, are characterized by left asymmetries; while large events resulting in more strain have symmetric shapes, similarly to the events at higher imposed strain rates $\dot{\epsilon}_a$~\cite{lehtinen2016glassy}. The rate dependence of the fitting parameters $\gamma$ and $a$ are shown in Fig.~\ref{fig:6}(b) as well as in Table~\ref{tab:table1}, including also the results for different system sizes, although only small size effects are observable. The rate dependence of the asymmetry parameter $a$ (right $y$-axis) shows that the imposed strain rates $\dot{\epsilon}_a=5000\,\mathrm{s}^{-1}$ and $\dot{\epsilon}_a=20000\,\mathrm{s}^{-1}$ are still characterized by a weak asymmetry, however higher rates appear to be symmetric up to a small error. The exponent $\gamma$ exhibits a similar rate dependence, namely $\gamma=1.25$ for the lowest imposed strain rate which then reaches the values $\gamma=1.5$ for higher rates. This value for high rates is in accordance with the exponent $\gamma=1.5\pm0.05$ obtained for the scaling of the average avalanche size with the duration [see inset of Fig.~\ref{fig:5}(a)], although here no clear rate dependence was observed. This discrepancy could be caused by the inability of Eq.~\eqref{eq:2} to accurately capture highly asymmetric avalanche shapes.
	
	\begin{figure}[h!]
		\centering
		\resizebox{0.95\columnwidth}{!}{\includegraphics{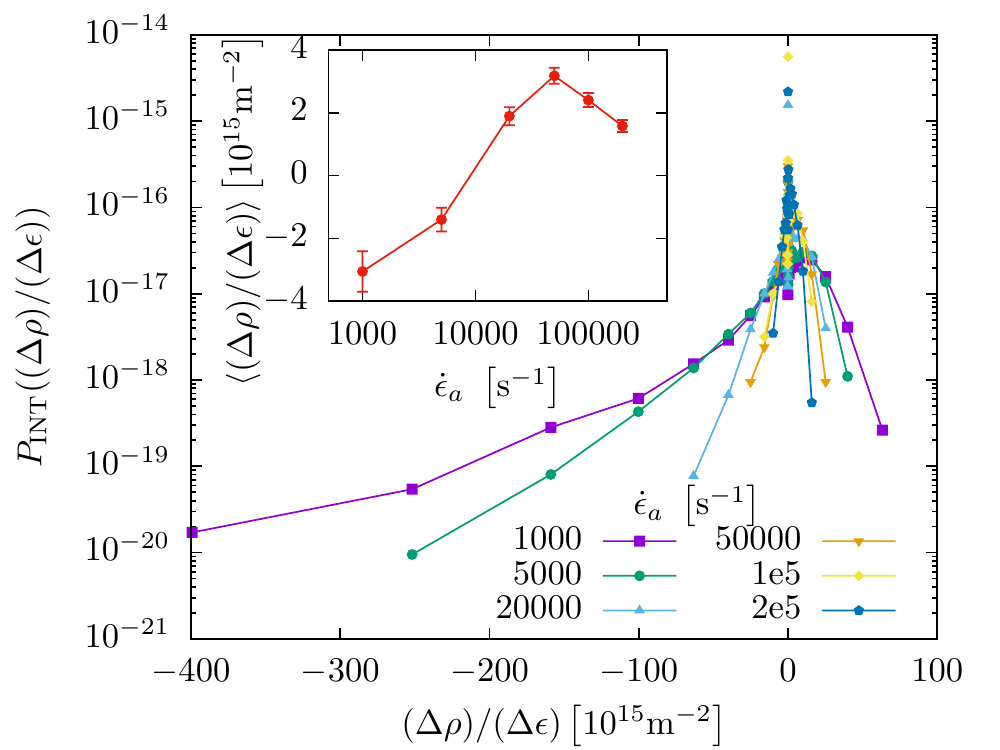}}
		\caption{Distributions of density differences divided by the corresponding strain increments $\Delta\rho$ between the ends and beginnings of the events in the power-law regime of the size and duration distributions divided by the corresponding strain increments $\Delta\epsilon$, for varying imposed strain rates $\dot{\epsilon}_a$, and for a system size with $N_0=20$ initial dislocation lines. The inset shows the average density difference divided by its strain increment as a function of the imposed strain rate $\dot{\epsilon}_a$.}
		\label{fig:7}
	\end{figure}

	Figure \ref{fig:7} shows the distributions of accumulated density differences scaled by the corresponding strain increments $(\Delta\rho)/(\Delta\epsilon)$ during events found in the power-law regime of the size and duration distributions. While for all the tested imposed strain rates $\dot{\epsilon}_a$ the scaled density differences of most events are close to zero, the distributions reveal the presence of events causing a significant decrease in the dislocation density, i.e. negative $(\Delta\rho)/(\Delta\epsilon)$. Moreover, the negative scaled density differences can be larger than the positive ones for the two lowest imposed strain rates $\dot{\epsilon}_a=1000\,\mathrm{s}^{-1}$ and $5000\,\mathrm{s}^{-1}$. To assess whether density decreases or increases dominate the avalanches, the inset shows the average scaled density differences as a function of the imposed strain rate. On average, for the lowest imposed strain rates, during the events in the power law regime the dislocation density decreases, however, as the loading rate is increased, the dislocation density keeps rising during the avalanches.
	
	Comparing this finding to the rate dependence of the dislocation density evolution [Fig.~\ref{fig:4}]  and the average avalanche shapes [Fig.~\ref{fig:6}], we can observe a transition from small loading rates, characteristic of individual temporally asymmetric avalanches and on average decreasing dislocation densities during the events, towards larger rates where the events are a consequence of several overlapping avalanches resulting in symmetric shapes and an increasing dislocation density. While similar findings have been presented for micropillar compression~\cite{PhysRevMaterials.3.080601,PhysRevLett.122.178001}, the presence of these two key physical regimes in the deformation of Al bulk single crystals extends our knowledge of plasticity, bringing us one step closer to optimizing the mechanical properties of materials by controlling the dislocation dynamics.

	\section{\label{sec:4} Conclusions}
	
	To summarize, our 3D DDD simulations on pure Al single crystals using a large number of random initial dislocation configurations to study the statistics of avalanches occurring during strain-controlled deformation process show that these are distributed by a power law with an exponential cutoff, where the exponent $\tau=1.0$ is independent of deformation rates and system sizes. This is different from the exponent $\tau=1.5$ observed for the same system using a quasistatic stress-controlled loading~\cite{lehtinen2016glassy}, as well as from the rate dependent exponent in simplified 2D DDD simulations~\cite{PhysRevE.104.025008}. In conclusion, while 2D DDD captures the special case of straight parallel edge dislocations moving in a single direction, the 3D representation of dislocations as flexible lines is recommended to model the plastic deformation process of realistic materials, such as the Al single crystals studied here.
	
	Contrary to the avalanche size and duration distributions, the average avalanche shapes exhibit a rate dependence where events at low deformation rates are characterized by shapes with left asymmetries. Meanwhile an increase of the strain rate shifts the events towards symmetric shapes. A transition from events reducing the dislocation density to events increasing it is observed at similar strain rates, highlighting the existence of two physical regimes characterized by different microscopic mechanisms within the same material. To identify, characterize and better understand these regimes a future in-depth theoretical as well as experimental study on the influence of material properties, including the crystal structure would be required~\cite{PhysRevLett.100.155502,biscari2016,alcala2020}. Additionally, studying avalanches due to strain-controlled loading in DDD simulations with quenched pinning interfering with dislocation motion would be of interest, given that there the dislocation avalanches have been found to exhibit depinning-like characteristics~\cite{ovaska2015quenched,PhysRevE.93.013309,salmenjoki2020plastic}.
	
	\begin{acknowledgments}
		The authors acknowledge the support of the Academy of Finland via the Academy Project COPLAST (project no. 322405). 
	\end{acknowledgments}
	
	\bibliographystyle{apsrev4-2}
	\bibliography{bibliography}% Produces the bibliography via BibTeX.
	
\end{document}